\documentclass[twocolumn]{aastex62}

\usepackage{comment}
\usepackage{multirow}

\newcommand{\unsim}{\mathord{\sim}}
\newcommand{\gaia}{\textit{Gaia}}
\newcommand{\galex}{\textit{GALEX}}

\newcommand{\Mjup}{$\mathrm{M_\mathrm{jup}}$}
\newcommand{\Msun}{$\mathrm{M_{\sun}}$}
\newcommand{\Rsun}{$\mathrm{R_{\sun}}$}

\graphicspath{{./}{figures/}}

\received{May, 2021}
\revised{}
\accepted{}
\submitjournal{ApJL}

\shorttitle{ZTFJ0038+2030}
\shortauthors{van Roestel et al.}
\begin{document}

\title{ZTFJ0038+2030: a long period eclipsing white dwarf and a substellar companion}

\correspondingauthor{Jan van Roestel}
\email{jvanroes@caltech.edu}

\author[0000-0002-2626-2872]{Jan van~Roestel}
\affiliation{Division of Physics, Mathematics, and Astronomy, California Institute of Technology, Pasadena, CA 91125, USA}

\author[0000-0002-6540-1484]{Thomas Kupfer}
\affiliation{Department of Physics and Astronomy, Texas Tech University, PO Box 41051, Lubbock, TX 79409, USA}

\author[0000-0002-0656-032X]{Keaton J.\ Bell}
\affiliation{DIRAC Institute, Department of Astronomy, University of Washington, Seattle, WA-98195, USA}
\affiliation{NSF Astronomy and Astrophysics Postdoctoral Fellow}

\author[0000-0002-7226-836X]{Kevin Burdge}
\affiliation{Division of Physics, Mathematics, and Astronomy, California Institute of Technology, Pasadena, CA 91125, USA}

\author[0000-0001-7016-1692]{Przemek Mr{\'o}z}
\affiliation{Division of Physics, Mathematics, and Astronomy, California Institute of Technology, Pasadena, CA 91125, USA}

\author{Thomas A. Prince}
\affiliation{Division of Physics, Mathematics, and Astronomy, California Institute of Technology, Pasadena, CA 91125, USA}

\author[0000-0001-8018-5348]{Eric C. Bellm}
\affiliation{DIRAC Institute, Department of Astronomy, University of Washington, 3910 15th Avenue NE, Seattle, WA 98195, USA}

\author{Andrew Drake}
\affiliation{Division of Physics, Mathematics, and Astronomy, California Institute of Technology, Pasadena, CA 91125, USA}

\author[0000-0002-5884-7867]{Richard Dekany}
\affiliation{Caltech Optical Observatories, California Institute of Technology, Pasadena, CA  91125}

\author[0000-0003-2242-0244]{Ashish~A.~Mahabal}
\affiliation{Division of Physics, Mathematics, and Astronomy, California Institute of Technology, Pasadena, CA 91125, USA}
\affiliation{Center for Data Driven Discovery, California Institute of Technology, Pasadena, CA 91125, USA}

\author{Michael Porter}
\affiliation{Caltech Optical Observatories, California Institute of Technology, Pasadena, CA  91125, USA}

\author{Reed Riddle}
\affiliation{Caltech Optical Observatories, California Institute of Technology, Pasadena, CA  91125, USA}

\author[0000-0002-1486-3582]{Kyung Min Shin}
\affiliation{California Institute of Technology, Pasadena, CA 91125, USA}

\author[0000-0003-4401-0430]{David L. Shupe}
\affiliation{IPAC, California Institute of Technology, 1200 E. California Blvd, Pasadena, CA 91125, USA}



\begin{abstract}
In a search for eclipsing white dwarfs using the Zwicky Transient Facility lightcurves, we identified a deep eclipsing white dwarf with a dark, substellar companion. The lack of an infrared excess and an orbital period of 10 hours made this a potential exoplanet candidate. We obtained high-speed photometry and radial velocity measurements to characterize the system. The white dwarf has a mass of $0.50\pm0.02$\,\Msun\ and a temperature of $10900\pm200$\,K. The companion has a mass of $0.059\pm0.004$\,\Msun\ and a small radius of $0.0783\pm0.0013$\,\Rsun. It is one of the smallest transiting brown dwarfs known and likely old, $\gtrsim 8$\,Gyr. The ZTF discovery efficiency of substellar objects transiting white dwarfs is limited by the number of epochs and as ZTF continues to collect data we expect to find more of these systems. This will allow us to measure period and mass distributions and allows us to understand the formation channels of white dwarfs with substellar companions.  
\end{abstract}



\section{Introduction} \label{sec:intro}
% WD BD
%\the\columnwidth
%\the\textwidth
%\the\textheight
Substellar objects mainly consist of hydrogen gas and are not massive enough to fuse hydrogen in their core ($M\lesssim M_\mathrm{HBL} \approx 0.07$\,\Msun; 73\,\Mjup). Substellar objects have masses in the range of $\unsim 0.3$--$73$\,\Mjup\ and are generally divided into two classes: brown dwarfs and giant exoplanets. There is no clear separation based on mass, but the distinction is based on the formation history (see \citealt{burrows2001} for an extended discussion). The formation of a brown dwarf is the same as that of more massive main-sequence stars: they are formed by gravitational instabilities in gas clouds and have elemental abundances similar to that of the interstellar medium. On the other hand, giant planets are formed by core accretion in a disk around a protostar, and have an enhanced metal abundance compared to the host star. Transit studies \citep[e.g.][]{carmichael2021}, radial velocity studies \citep[e.g.][]{shahaf2019}, and microlensing observations \citep[e.g.][]{han2016,poleski2017} of main-sequence stars show that substellar objects exist in orbits of $\unsim$ 1\,AU. The mass distribution suggests that there are two distinct populations, giant planets with masses $\lesssim30$\,\Mjup, and brown dwarfs with masses $\gtrsim 60$\,\Mjup, with a gap in between, the brown dwarf desert \citep{marcy2000,grether2006,sahlmann2011,ma2014}. However, recent discoveries by TESS shows that substellar objects do span the entire mass-range \citep{carmichael2020}.

At the end of their life, main sequence stars go through a red-giant phase, which significantly affects nearby substellar companions \citep[see][ for giant planets around red giants]{grunblatt2018}.
If a substellar companion is in a close enough orbit ($\approx 200$--$1000$\,\Rsun, 1--5\,AU for RGB and AGB giants), the system forms a common-envelope \citep{ivanova2013}. While more massive objects (brown dwarfs and low mass red dwarfs) are known to survive this process and end up as a short period binary \citep{casewell2018,casewell2020,casewell2020a}, lower mass substellar companions are predicted to merge during a common-envelope event or get evaporated as soon as the hot white dwarf emerges from the common envelope \citep{soker1998,nelemans1998,bear2011}. However, recent work \citep{lagos2021} suggest that giant planet can survive common envelope evolution. 
There are also alternative pathways to form short-period white dwarfs--planets systems, including the formation of second-generation planets from gas around the white dwarf \citep[e.g.][]{perets2010}, or capture and/or inward migration of distant planets \citep[e.g.][]{stephan2020}.

There are many indications that short-period white dwarf--planet systems do exist: dust discs have been observed around white dwarfs \citep[e.g.][]{zuckerman1987}; the detection of heavily polluted white dwarfs \citep{koester2014}; transiting debris around WD 1145+017 \citep{vanderburg2015} and other white dwarfs \citep{vanderbosch2020,guidry2020}; and
white dwarf WD J0914+1914 which shows accreting material of a Neptune-like composition \citep{gansicke2019}.

There have been many searches for exoplanets transiting white dwarfs, e.g. \citet{faedi2011,fulton2014,vansluijs2018,dame2019,rowan2019}, but all of them did not find any candidates.
In addition, \citet{vangrootel2021} searched for exoplanets around bright SdB stars, but were also only able to put an upper limit to the planet occurrence rate.
\citet{vanderburg2020} discovered a white dwarf with a substellar companion with an orbital period of 1.4 days (WD J0914+1914). After carefully analysing optical and infrared lightcurves of the grazing eclipse, they conclude that the companion is a giant planet with a mass of $\lesssim14$\Mjup. 

The discovery of the giant planet candidate around WD J0914+1914, which is bright and nearby, suggests that there are more eclipsing white dwarfs with (low mass) substellar companions. Because the white dwarf is small and hot, eclipses of substellar companions around white dwarfs are deep, and eclipses are the best method to detect these systems \citep[e.g.][]{bell2020,agol2011a}. However, eclipses alone are not sufficient to determine the nature of the object. The radius of substellar objects is almost invariant of their mass for masses between 1 Jupiter mass and 0.07\Msun brown dwarfs \citep{hatzes2015}, which means the optical eclipse is identical for a Jupiter mass-object and a brown dwarf. However, the tidal disruption period is a function of density \citep{rappaport2013,rappaport2021}, which means that we can determine (using mass-radius models) what the \textit{minimum} mass is. High mass substellar objects have a high density and can exist in orbital periods as short as 80 minutes, but low mass, low-density Jupiter-like objects exceed their Roche radius at periods as long as 9 hours and can only exist at longer periods.

%A detection of the companion using infrared observations is need \citep{vanderburg2020}

%However, giant planets have a low density which means that tidal forces destroy them if they are too close to the white dwarf. For massive giant planets (\unsim30\Mjup), this occurs at orbital periods of $unsim2$\,h, while for a 1\Mjup planet, this happens at an orbital period of $unsim15$\,h. This explains why these systems have not been found yet; at long orbital periods, both the eclipse probability is low and the eclipse duty-cycle short, $\lesssim 1\%$. Current and upcoming surveys like TESS, ZTF, ATLAS, and \LSST will collect enough epochs to find more of these systems \citep{agol2011,bell2020}. However, substellar companions ranging from 1 to 80 \Mjup (brown dwarfs) have almost identical sizes, which means their transit appears identical. 

In this paper, we present the discovery and characterization of ZTFJ003855.0+203025.5 (ZTFJ0038+2030), an eclipsing white dwarf with a substellar companion with an orbital period of 10 hours.
To identify the eclipses, we used the Zwicky Transient Facility lightcurves \citep{graham2019,bellm2019,masci2019,dekany2020}. The system was identified in a search for deep eclipsing white dwarfs. We searched the combined PSF-photometry and alert photometry lightcurves of white dwarfs \citep{Gentilefusillo2019} for deep eclipses and identified the period using the BLS algoritm \citep{Kovacs2002}. For more details, see \citet{vanroestel2021} and Van Roestel (2021b) (in prep). ZTFJ0038+2030 showed a complete eclipse in the $g$ and $r$ band and the eclipse duration is short, consistent with that of a substellar object. It also showed no excess luminosity in the \gaia\ HR-diagram and no infrared excess in Pan-STARRS colors. Because of these properties, we prioritised it for followup observations to determine the nature of the companion.

We obtained followup photometry and spectroscopy (Section \ref{sec:followup}), which we used the characterize the system (Section \ref{sec:analysis}). We present the mass, radius, and temperature measurements in Section \ref{sec:results}. We compare this binary system with other white dwarfs with substellar companions, and discuss the implications of this discovery for future searches for giant exoplanets around white dwarfs with ZTF. We end with a summary.

\begin{table}[]
        \caption{Brightness of ZTFJ0038+2030 in different bands. Gaia eDR3 data was used \citep{brown2020a}, with the geometric distance from \citet{bailer-jones2021}. `$^V$' indicates that the magnitudes are in the Vega system, other magnitudes are in the AB-system.}
    \centering
   \begin{tabular}{l|l}
    \hline
    \hline
        RA &  $00^{\rm h}$38$^{\rm m}$55.0$^{\rm s}$ \\ 
        Dec & $20^{\circ}$30$^{'}$26.1$^{''}$ \\
        $G^V$ & 	17.70 \\
        $BP^V$ & $17.76 \pm 0.01$ \\
        $RP^V$ & $17.63 \pm 0.01$ \\
        parallax  & $7.19\pm0.11$\,mas\\
        distance  & $138.3^{+1.7}_{-1.9}$\,pc \\
        \galex\ FUV & $20.37 \pm 0.24$ \\
        \galex\ NUV & $18.58 \pm 0.06$  \\
        ZTF-$g$ & $17.70 \pm 0.02$\\
        ZTF-$r$ & $17.78 \pm 0.03$\\
        ZTF-$i$ & $17.95 \pm 0.03$\\
        PS-$g$ & $17.705 \pm 0.005$ \\
        PS-$r$ & $17.786 \pm 0.002$\\
        PS-$i$ & $17.931 \pm 0.006$\\
        PS-$z$ & $18.093 \pm 0.005$\\
        PS-$y$ & $18.18 \pm 0.02$\\
        WISE-$W1^V$ & $17.87 \pm 0.11$ \\
        WISE-$W2^V$ & $17.63 \pm 0.29$\\
    \end{tabular}
    \label{tab:overview}
\end{table}

\section{Followup data}\label{sec:followup}

\begin{table*}
 \centering
 \caption{Summary of the followup observations}
  \begin{tabular}{lclrrc}
  \hline\hline
Date &    UT  &  Tele./Inst. & N$_{\rm exp}$ & Exp. time (s) & Wavelength \\
  \hline
2020-07-15  & 11:22 - 11:57 & P200/CHIMERA & 400 & 5.0 & $g$ \\
2020-07-15  & 11:22 - 11:57 & P200/CHIMERA & 400 & 5.0 & $z$ \\
2020-07-21  & 12:18 - 12:39 &  Keck/ESI/Echellete & 2 & 600\phantom{.0} &  4000 - 10000\AA \\
2020-09-12  & 12:17 - 12:38 &  Keck/ESI/Echellete & 2 & 600\phantom{.0} &  4000 - 10000\AA \\
2020-09-12  & 14:16 - 14:37 &  Keck/ESI/Echellete & 3 & 600\phantom{.0} &  4000 - 10000\AA \\
   \hline
\end{tabular}
\label{tab:observ}
\end{table*}

\subsection{CHIMERA fast cadence photometry}
 We obtained high-speed photometry in the $g$ and $z$ filters using CHIMERA (see Table \ref{tab:observ}).
CHIMERA \citep{harding2016} is a dual-channel photometer that uses frame-transfer, electron-multiplying CCDs mounted on the Hale 200-inch (5.1 m) Telescope at Palomar Observatory (CA, USA). The pixelscale is 0.28 arcsec/pixel (unbinned). We used the conventional amplifier and used 2x2 binning on most nights to reduce the readout noise. Each of the images was bias subtracted and divided by twilight flat fields. We used the ULTRACAM pipeline to do aperture photometry \citep{dhillon2007}. We used an optimal extraction method with a variable aperture of 1.5 the FWHM of the seeing (as measured from the reference star). A differential lightcurve was created by simply dividing the counts of the target by the counts from the reference star. Timestamps of the images were determined using a GPS receiver.

\subsection{ESI}
We used the Echellete Spectrograph and Imager \citep[ESI, ][]{sheinis2002} mounted at KeckII to obtain medium-resolution spectra ($R\approx6000$). CuAr arc exposures were taken at the beginning of the night. The spectra were reduced using the \textit{MAKEE}\footnote{http://www.astro.caltech.edu/$\sim$tb/ipac$\_$staff/tab/makee/} pipeline following the standard procedure: bias subtraction, flat fielding, sky subtraction, order extraction, and wavelength calibration.

\subsection{Archival photometry}
To be able to study the spectral energy distribution, we obtained photometry data from multiple other survey telescopes (see Table \ref{tab:overview}):  \gaia\ eDR3 \citep{GaiaeDR3}, \textit{Galex} \citep{GALEX2017}, Pan-STARRS \citep{PS1}, and \textit{WISE} \citep{CatWISE2020}. 
We used zero points for each of the filters to convert the magnitudes to a flux.

\section{Analysis}\label{sec:analysis}

\subsection{Ephemeris}\label{subsec:ephemeris}
We determine the ephemeris by measuring the mid-eclipse time from the CHIMERA $g$ lightcurve. We then use the best model from the Chimera $g$ data and use it fit all ZTF data. In addition, we noticed that there is one non-detection on 2012-11-1 in Palomar Transient Factory data (out of 94 observations). We add this epoch with half the eclipse duration as uncertainty as a prior ($BJD_{TDB}=2456232.8854\pm0.0018$). This results in an ephemeris of:
\begin{equation}
BJD(TDB) = 2459045.985194(2) + 0.431\,920\,8 (14)
\end{equation}

\subsection{Spectra and radial velocity amplitude}\label{subsec:RV}
The spectra show a typical DA white dwarf spectrum with broad Balmer absorption lines. No features from the brown dwarf can be seen, including any Balmer emission due to irradiation \citep[e.g.][]{parsons2018}.

Radial velocities of the ESI spectra were measured by fitting a Gaussians, Lorentzians, and polynomials to the hydrogen lines to cover the continuum, line, and line core of the individual lines using the \textit{FITSB2} routine \citep{napiwotzki2004}. The procedure is described in full detail in \citet{kupfer2020,kupfer2017,kupfer2017a}. We fitted the wavelength shifts compared to the rest wavelengths using a $\chi^2$-minimization.

\begin{figure*}
\centering
\includegraphics{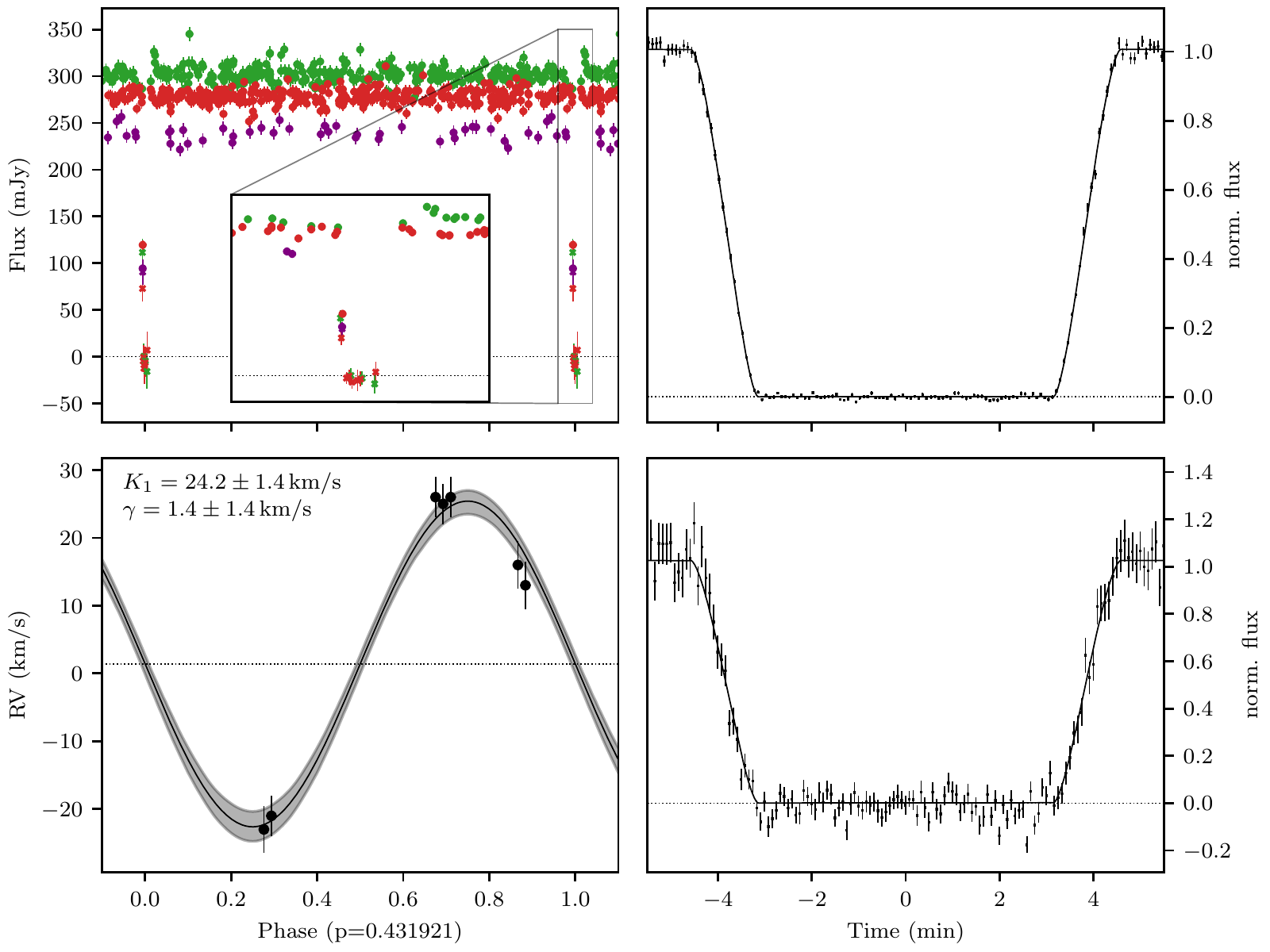}
\caption{The top left panel shows the ZTF $gri$ data (green, red, purple) folded to the period. The bottom left panel shows the ESI radial velocity measurements. The best-fit radial velocity curve is overplotted, with in grey the the 1-standard deviation, and the dotted line indicates the systematic velocity. The right two panels show the CHIMERA $g$ (top) and $z$ data (bottom) with the best-fit \textit{lcurve} model overplotted.}
\label{fig:alldata}
\end{figure*}

To determine the radial velocity semi-amplitude of the white dwarf ($K_1$), we fit the radial velocity measurements using a sinusoid with a fixed period and zero phase based on the ephemeris determined from the ZTF data. The two remaining free parameters are the amplitude ($K_1$) and a systematic velocity ($\gamma$). We use the \textit{emcee} \citet{ForemanMackey2013} to determine the best value and uncertainty:
$K_1 = 24.2 \pm 1.4$\,km/s (Fig. \ref{fig:alldata}).

\subsection{Spectral energy distribution}\label{subsec:SEDmodelling}
\begin{figure*}
    \centering
    \includegraphics{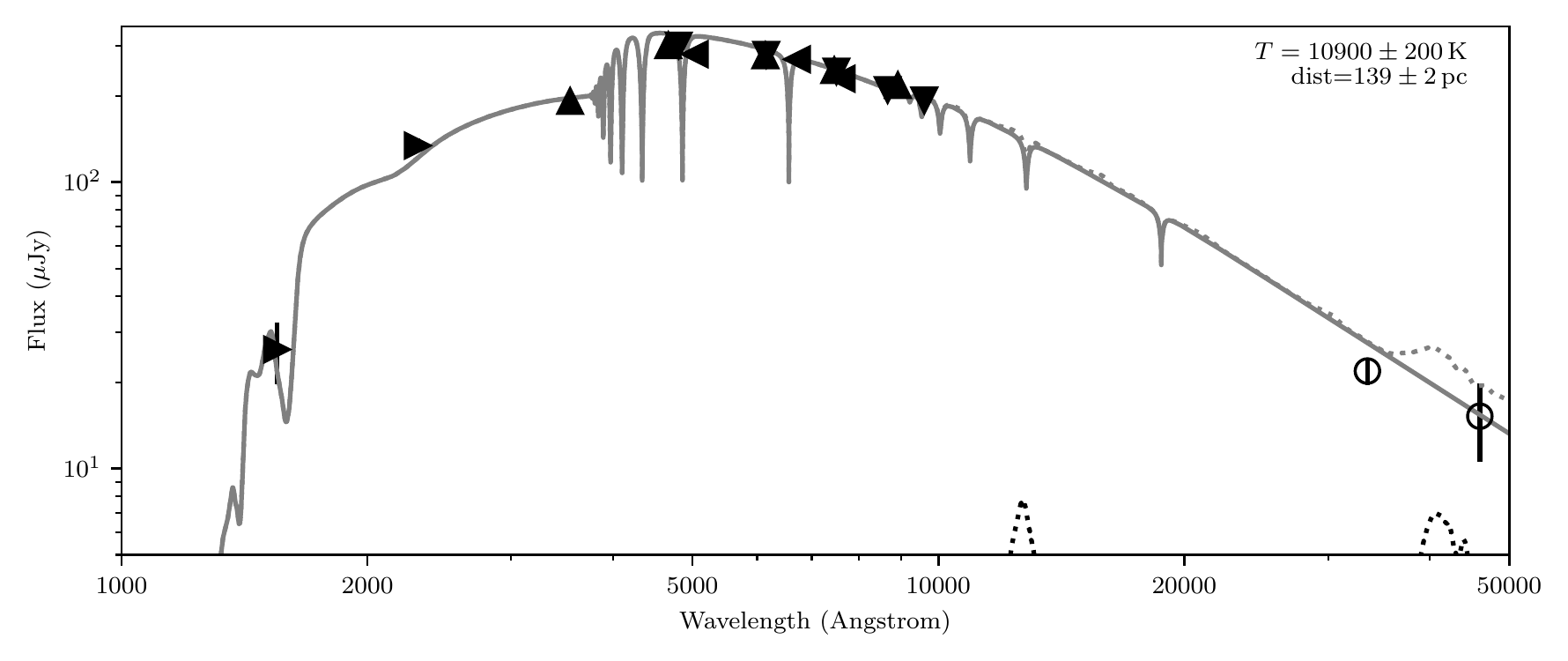}
    \caption{The spectral energy distribution of the system. Markers show \textit{Galex}, \gaia\ DR3, Pan-STARRS, median ZTF $gri$, and \textit{WISE} data. Open markers indicate data not used to constrain the fit. The grey solid line shows the best-fit DA white dwarf model. The dotted line shows the SED of the best-fit DA model with a 900\,K brown dwarf model added.}
    \label{fig:SED}
\end{figure*}

To determine the white dwarf temperature, we fit the observed spectral energy distribution with white dwarf models (see Fig. \ref{fig:SED}). We use a grid of DA white dwarf models by \citet{koester2009} and use bilinear interpolation to be able to generate a model for any temperature and surface gravity value. We use the extinction law by \citet{fitzpatrick1999} to account for any dust extinction.
To compare the model spectra with the data, we convolve the model spectra with the filter response curves\footnote{\url{http://svo2.cab.inta-csic.es/theory/fps/}} \citep{rodrigo2012,rodrigo2020}. We use Gaussian priors on the parallax using the \gaia\ eDR3 data, the radius estimate from the lightcurve, and an $E_{BV}$ value from Pan-STARRS extinction estimates \citep{green2018}. We again use \textit{emcee} to estimate the best-fit values and uncertainties. Using this method, we estimate the white dwarf temperature to be: $T_\mathrm{WD}=10900\pm200$\,K (Fig. \ref{fig:SED}).

\subsection{Lightcurve modelling}\label{subsec:lcmodelling}
\label{sec:lcfit}

We modelled the high cadence lightcurves using the package \textit{ellc} \citep{maxted2016}.
We use a spherical star to model the white dwarf, and use a Roche-lobe geometry for the companion. The free parameters for this model are the mid-eclipse time ($t_0$), inclination ($i$), mass-ratio ($q$), the radii divided by the semi-major axis of both objects ($r_{1,2} \equiv R_{1,2}/a$), the semi-major axis ($a$), and the surface brightness ratio ($J_{g,z}$).

%., and the velocity scale ($(K_1+K_2)/\sin i$). 

We used a number of fixed parameters in the binary model. First, we use the orbital period obtained from the ZTF data (Section \ref{subsec:ephemeris}). For limb-darkening of the white dwarf we the values for T=$10\,000$\,K and $\log(g)=8.0$ as calculated by \citep{claret2020}.

In addition, we imposed two restrictions on the white dwarfs. The first is that it cannot be smaller than a zero-temperature white dwarf. The second constrain is a Gaussian prior on the white dwarf radius relative to the white dwarf M-R relation with an uncertainty of 5\%. We use the approximation of the mass-radius relation of Eggleton from \citet{Rappaport1989}. As a final constraint, we use a Gaussian prior on the radial velocity amplitude ($K_1$) of the white dwarf (see Section \ref{subsec:RV}). 

To find most probable parameter values and uncertainties, we again use \textit{emcee}.

\section{Results}\label{sec:results}
\begin{table}\label{tab:results}
\centering
\caption{The binary parameters are determined by modelling the lightcurves using \textit{ellc}. The top section lists model parameters, the bottom part shows derived parameters. We fixed the orbital period ($^f$), and for the radius of the white dwarf ($R_1$) and radial velocity amplitude ($K_1$) we used a prior ($^p$). We use the 95\% percentile to determine upperlimits.}
\label{tab:pars}
\begin{tabular}{l|l}
\hline
$p^f$ (d)      & 0.431\,920\,8 (14) \\
$t_0$ ($\mathrm{BJD_{TBD}}$) & $2459045.985194(2)$  \\
$q$ & $0.1167^{+0.0075}_{-0.0068}$ \\
$i$ ($^{\circ}$) & $89.71^{+0.12}_{-0.13}$ \\
$r_1$ &  $0.007195^{+0.000075}_{-0.000078}$ \\
$r_2$ & $0.03934^{+0.00033}_{-0.00019}$ \\
$a$ (\Rsun) & $1.987^{+0.030}_{-0.022}$ \\
$J_g$ & $\lesssim{0.000035}$  \\
$J_z$ & $\lesssim{0.00014}$  \\
\hline
$M_1$ (\Msun) & $0.505^{+0.024}_{-0.018}$ \\
$M_2$ (\Msun) & $0.0593^{+0.0036}_{-0.0039}$  \\
$R_1^p$ (\Rsun) & $0.01429^{+0.00022}_{-0.00017}$ \\
$R_2$ (\Rsun) & $0.0783^{+0.0013}_{-0.0011}$ \\
$\log(g_1)$ (cgs) & $7.832^{+0.013}_{-0.013}$  \\
$\log(g_2)$ (cgs) & $5.425^{+0.02}_{-0.03}$ \\
$K_1^p$ (km/s) & $24.4^{+1.4}_{-1.4}$ \\
$K_2$ (km/s) & $208.4^{+3.7}_{-2.9}$ \\
 $\overline{\rho}_2$  ($\mathrm{g/cm^3}$) &  $174^{+9}_{-11}$ \\
\end{tabular}
\end{table}

We measured the binary properties by analysing the spectral energy distribution, ZTF lightcurves, phase-resolved spectroscopy, and high cadence $g$- and $z$-band lightcurves. The results are summarized in Table \ref{tab:results} and the posterior of the lightcurve modelling is shown in the appendix (Fig. \ref{fig:corner}). 

The mass of the companion, which is mostly set by the radial velocity semi-amplitude measurement, is $M_2=0.0593\pm0.004$\Msun, and a radius of $R_2=0.0783^{\phantom{-}0.0013}_{-0.0011}$\Rsun. This is consistent with a brown dwarf. Using the $z$-band surface brightness ratio, we estimate that the temperature of the brown dwarf is $\lesssim 1550$\,K.

The mass of the white dwarf is $0.50\pm0.02$\,\Msun, which is typical for a white dwarf \citep{kepler2007}. The white dwarf radius ($R_1=0.01429 \pm 0.00020$\,\Rsun) is consistent with the white dwarf M-R relation, which is what we enforced with a prior. The temperature of the white dwarf is $T_1=10900\pm200$\,K. 

The orbital separation of the binary system is $a=1.987 \pm 0.027$\Rsun\ and the inclination of this system is $i=89.71 \pm {0.13} ^\circ$. The corresponding impact parameter is $b\lesssim0.18$. %The probability for this system to show a full eclipse is 3.1\%.
% limiting inclination = 88.1

\section{Discussion}
\subsection{The nature of the substellar companion}
\begin{figure*}
    \centering
    \includegraphics{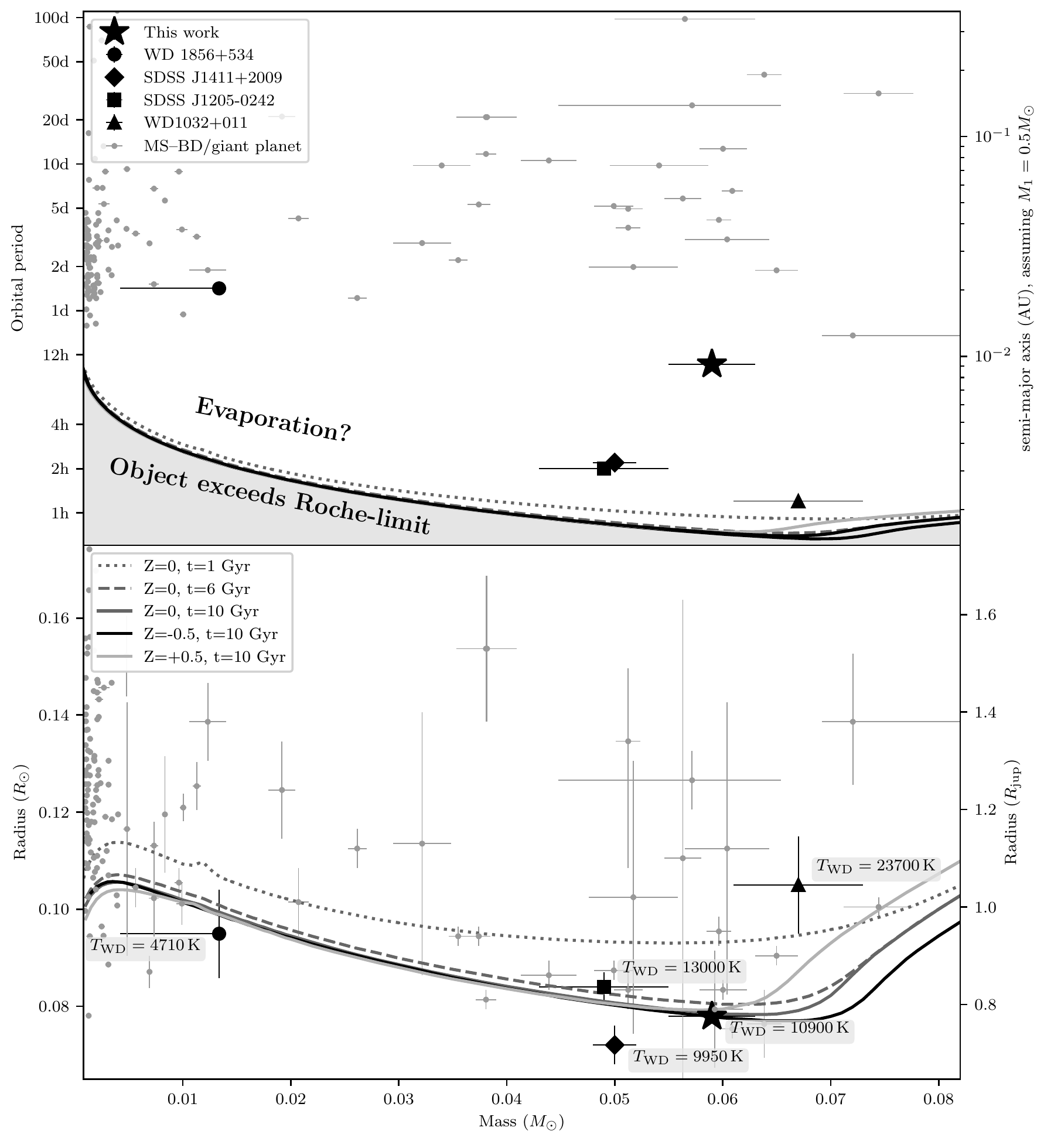}
    \caption{The characteristics of substellar companions that are eclipsing white dwarfs (black markers). The parameters of the other eclipsing white dwarfs with substellar companions are taken from \citet{vanderburg2020}, \citet{casewell2020}, \citet{parsons2018}, \citet{littlefair2014}. Grey points show brown dwarfs and giant planets around other stellar types taken \citet{carmichael2021} and \citet{chen2017a}. The top panel shows the orbital period versus the mass. The lower-left region is off-limits as the object would exceeds the Roche-limit \citep{rappaport2013}. Low-mass objects close to hot white dwarfs are also affected by photo-evaporation \citep{soker1998,nelemans2004,bear2011}. The bottom panel shows the radius versus the mass. The white dwarf temperature is indicated next to each marker. Models are taken from \citet{marley2018}. Uncertainties for low mass objects are omitted for clarity. }
    \label{fig:MR}
\end{figure*}

 In Fig.\ \ref{fig:MR}, we plot the mass and radius of the companion, and compare it to models by \citet{marley2018}. The measured mass and radius agree with models of 10 Gyr old brown dwarfs with $Z\gtrsim0$ abundances. The models predict a temperature of $\unsim 800$--$900$\,K, which is well below the upper limit we determined from the $z$-band surface brightness ratio. In Fig.\ \ref{fig:SED} we plot the best fit white dwarf spectrum with the spectrum of a 900\,K brown dwarf added. Such a low temperature is consistent with the lack of any excess emission in the \textit{WISE} bands. If we assume a solar abundance or lower, the age of the brown dwarf (and therefore the system) is $\gtrsim8$\,Gyr. 
 %In conclusion, the substellar companion is an old brown dwarf. 

 Compared to other substellar objects that are eclipsing white dwarfs, the mass and radius do not stand out and are similar to other brown dwarfs. This object does stand out because of its orbital period, which at 10 hours is an order of magnitude larger than the three other known brown dwarfs orbiting white dwarfs. This means that the amount of irradiation by the white dwarf is relatively low. Using a simple blackbody approximation \citep{littlefair2014}, we estimate that the temperature of the brown dwarf is only increased by $\unsim50$\,K due to irradiation by the white dwarf. This fact, and the systems relative brightness, make this system a good prototype system for long period white dwarf brown dwarf systems. 
 
 %If we compared the system with all eclipsing brown dwarfs, we see that the brown dwarf  
 
 %This fact, combined with the relative brightness the systems, arguably makes ZTFJ0038+20 the best currently known system to test the mass-radius models of brown dwarfs using high precision radial velocity measurements and lightcurves.  

\subsection{Formation history}
%Since the companion is a brown dwarf and not a giant planet, standard common envelope evolution can explain this system. Given the mass of the white dwarf, it could have a He core (0.47\Msun, \citealt{han2003}) or a CO core \citep[e.g.][]{marigo2013}, see also \citet{parsons2017a}. The first would have been formed after a common envelope in the RGB phase, while the second is the result of a common envelope during the AGB phase. We note the He-WD might have emerged the common envelope as a SdB-star and appeared as a HW Vir system (an eclipsing SdB-dM/BD system, \citealt{schaffenroth2019}).

Since the companion is a brown dwarf and not a giant planet, standard common envelope evolution can explain the formation of this system. Given that the mass the white dwarf is $\geq 0.47$\,M$_\odot$\, the white dwarf has very likely a CO core (see e.g. \citealt{marigo2013} and also \citealt{parsons2017a} for observational evidence) which allows for two formation scenarios. In the first scenario, the white dwarf could have formed during a common envelope phase on the Asymptotic Giant branch (AGB) after helium core exhaustion. The second scenario is that the common envelope happened at the tip of the Red giant branch (RGB), just after the helium flash \citep{han2003} which would result in a white dwarf with a mass close to 0.47\,\Msun.
%Because the mass of the white dwarf is close to the helium core mass for a helium flash, the common envelope could have also happened at the tip of the Red giant branch (RGB) just after the helium flash \citep{han2003}. 
In that scenario the white dwarf would have after the common envelope evolved into a hot subdwarf (sdB) and appeared as an HW Vir system before it evolved into a white dwarf with a brown dwarf companion after helium exhaustion in the sdB. Several sdB + brown dwarf systems are known, although typically seen with shorter orbital periods \citep[e.g.][]{geier2011, schaffenroth2015,schaffenroth2018,schaffenroth2019}.

The initial to final mass relation for the white dwarfs suggests that the white dwarf progenitor was approximately a $1$--$2$\,\Msun\ main-sequence star. This corresponds to a main-sequence lifetime of $10$--$2$\,Gyr \citep{catalan2008,marigo2013,cummings2018}. The cooling age of the white dwarf is approximately $\unsim 400$\,Myr \citep{koester2009}. This is consistent with the age estimate based on the brown dwarf radius. 

The white dwarf will slowly cool down and the period will slowly decrease due to gravitational wave radiation. It will take $\unsim$135\,Gyr to reach an orbital period of $\unsim$40 minutes \citep{rappaport2021}, at which point the white dwarf will be $\unsim$1000 K. Roche-lobe overflow will commence and system becomes a cataclysmic variable \citep{littlefair2003}. The accretion flow will heat up the white dwarf again while the period increases. This will slowly drain the brown dwarf and the system ends up as a `period-bouncer'; a very low accretion rate CV with an orbital period of $\approx$90 minutes \citep[e.g][]{pala2018}.

\subsection{Implications for searches for giant planets transiting white dwarfs with ZTF}
Here, we briefly discuss the detection efficiency of our search and the occurrence rate of white dwarfs with transiting substellar objects. A detailed simulation is beyond the scope of this paper and we limit ourselves to an order of magnitude estimate only.

To find ZTFJ0038+2030 we searched the ZTF lightcurves of the \gaia\ white dwarf catalog by \citet{Gentilefusillo2019} which contains 486\,641 candidate white dwarf over the entire sky. There are 129\,148 white dwarf brighter than 20 mag with more than 80 epochs in their ZTF lightcurve. Based on the number of epochs in these lightcurves, we estimate an average recovery efficiency of 15-25\% (simply the probability of getting 7-5 in-eclipse points). We note that we recovered the other three eclipsing WD-BD systems in Fig. \ref{fig:MR}.

With the discovery of ZTFJ0038+2030 and the discovery by \citet{vanderburg2020}, there are now two known long-period ($\gtrsim 10$\,h) transiting substellar objects around a white dwarfs; the first most likely a giant planet and the second most likely a brown dwarf. This suggests that, at longer orbital periods, the occurrence rate is the same order of magnitude. More systems need to be found and characterized in order to determine the mass distribution and determine which of the formation channels are important in the formation of these objects.

Currently, the ZTF detection efficiency is limited by the number of epochs available per white dwarf. As more epochs are obtained, ZTF will be able to identify narrower eclipses, which means that longer period systems can be identified. Based on the recovery efficiency of ZTFJ0038+2030 we estimate that ZTF will find another $3-6$ similar objects as it keeps on accumulating more data. Other surveys like \gaia, ATLAS, and BlackGEM can be used to find similar systems over the whole sky. In the near future, the Vera C. Rubin observatory \citep{ive2019} will find many more white dwarfs with exoplanets, possibly down to earth-sized objects \citep{agol2011}.

\section{Summary and conclusion}
Using ZTF photometry, and \gaia\ and Pan-STARRS data, we discovered an eclipsing binary composed of a white dwarf and a substellar companion with an orbital period of 10 hours. 
We used follow-up photometry and spectroscopy to measure the binary parameters. This shows that the substellar companion is an $\gtrsim8$Gyr old brown dwarf with a mass of 0.06\Msun, and the white dwarf a 0.50\Msun, CO white dwarf. The system is relatively bright, and a good prototype system where the brown dwarf suffers minimal irradiation. It is also a useful target for eclipse timing to find circum-binary objects \citep[e.g. NNser, ][]{marsh2014} as brown dwarf are not expected to show eclipse time variations due to Applegate's mechanism \citep{applegate1992,bours2016}.

\acknowledgments
JvR  is  partially  supported  by  NASA-LISA  grant 76280NSSC19K0325.
K.J.B. is supported by the National Science Foundation under Award AST-1903828.

We thank Lars Bildsten for trading one hour of ESI-Keck time.

Based on observations obtained with the Samuel Oschin Telescope 48-inch and the 60-inch Telescope at the Palomar Observatory as part of the Zwicky Transient Facility project. ZTF is supported by the National Science Foundation under Grant No. AST-1440341 and a collaboration including Caltech, IPAC, the Weizmann Institute for Science, the Oskar Klein Center at Stockholm University, the University of Maryland, the University of Washington, Deutsches Elektronen-Synchrotron and Humboldt University, Los Alamos National Laboratories, the TANGO Consortium of Taiwan, the University of Wisconsin at Milwaukee, and Lawrence Berkeley National Laboratories. Operations are conducted by COO, IPAC, and UW.

Based on observations obtained with the 200-inch Hale Telescope at the Palomar Observatory as part of the Zwicky Transient Facility project. The Hale telescope is operated by the Caltech Optical Observatories.

Some of the data presented herein were obtained at the W.M. Keck Observatory, which is operated as a scientific partnership among the California Institute of Technology, the University of California and the National Aeronautics and Space Administration. The Observatory was made possible by the generous financial support of the W.M. Keck Foundation. The authors wish to recognize and acknowledge the very significant cultural role and reverence that the summit of Mauna Kea has always had within the indigenous Hawaiian community. We are most fortunate to have the opportunity to conduct observations from this mountain.

This research has made use of the VizieR catalogue access tool, CDS, Strasbourg, France. This research made use of NumPy \citep{harris2020array}; matplotlib, a Python library for publication quality graphics \citep{Hunter:2007}; SciPy \citep{Virtanen_2020}; Astropy, a community-developed core Python package for Astronomy \citep{2018AJ....156..123A, 2013A&A...558A..33A}. 

This work has made use of data from the European Space Agency (ESA) mission {\it Gaia} (\url{https://www.cosmos.esa.int/gaia}), processed by the {\it Gaia} Data Processing and Analysis Consortium (DPAC, \url{https://www.cosmos.esa.int/web/gaia/dpac/consortium}). Funding for the DPAC has been provided by national institutions, in particular the institutions participating in the {\it Gaia} Multilateral Agreement.  The Pan-STARRS1 Surveys (PS1) have been made possible through contributions of the Institute for Astronomy, the University of Hawaii, the Pan-STARRS Project Office, the Max-Planck Society and its participating institutes, the Max Planck Institute for Astronomy, Heidelberg and the Max Planck Institute for Extraterrestrial Physics, Garching, The Johns Hopkins University, Durham University, the University of Edinburgh, Queen's University Belfast, the Harvard-Smithsonian Center for Astrophysics, the Las Cumbres Observatory Global Telescope Network Incorporated, the National Central University of Taiwan, the Space Telescope Science Institute, the National Aeronautics and Space Administration under Grant No. NNX08AR22G issued through the Planetary Science Division of the NASA Science Mission Directorate, the National Science Foundation under Grant No. AST-1238877, the University of Maryland, and Eotvos Lorand University (ELTE). 


\vspace{5mm}
\facilities{P48(ZTF), P200:5.0m (CHIMERA), Keck2:10m (ESI),}


\software{astropy \citep{astropycollaboration2013}, 
          Makee, 
          ellc
          \citep{maxted2014a}, 
          scipy, 
          emcee \citep{ForemanMackey2013}
          }



% APPENDIX
\appendix
\section{Appendix information}

\begin{figure*}
    \centering
    \includegraphics[width=\textwidth]{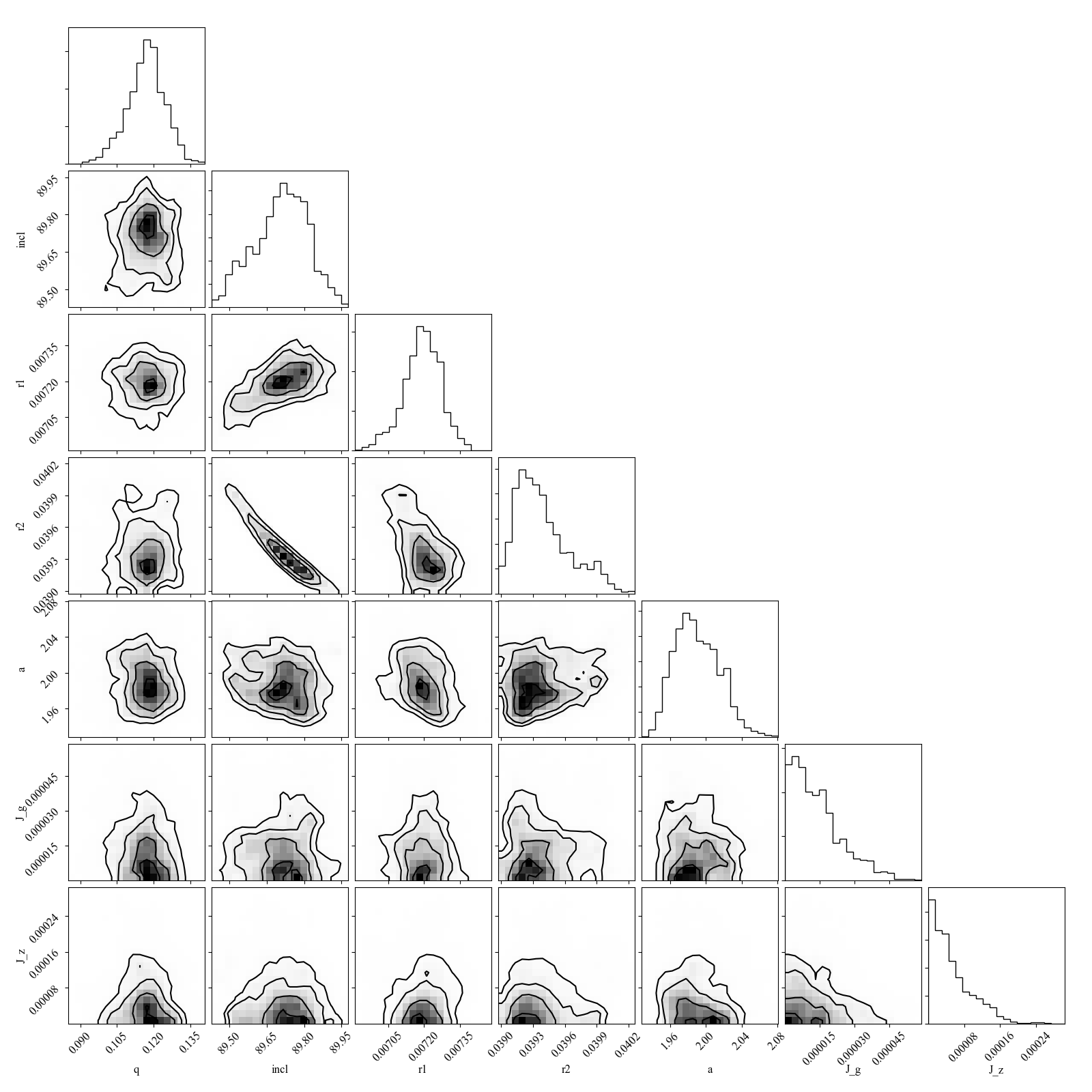}
    \caption{The posterior distribution of the fit to the lightcurve data using \textit{ellc} and \textit{emcee}.}
    \label{fig:corner}
\end{figure*}




\bibliography{references,references2}
\bibliographystyle{aasjournal}



\end{document}